\documentclass[a4paper]{article}

\usepackage{INTERSPEECH2019}
\usepackage{float}
\title{Margin Matters: Towards More Discriminative Deep Neural Network Embeddings for Speaker Recognition}
\name{Xu Xiang$^{1*}$, Shuai Wang$^1$, Houjun Huang$^2$, Yanmin Qian$^1$, Kai Yu$^1$\thanks{*Work done during an internship at AISpeech Co., Ltd., China}\thanks{This work has been supported by the National Key Research and Development Program of China (Grant No.2017YFB1002102), the China NSFC project (No. U1736202) and the Jiangsu NSFC project (No. BE2016078)}}
\address{
  $^1$MoE Key Lab of Artificial Intelligence \\
SpeechLab, Department of Computer Science and Engineering \\ 
Shanghai Jiao Tong University, Shanghai, China \\
  $^2$AISpeech Co., Ltd., China}
\email{\{chinoiserie, feixiang121976, yanminqian, kai.yu\}@sjtu.edu.cn, houjun.huang@aispeech.com}

\begin{document}

\maketitle

\begin{abstract}
  Recently, speaker embeddings extracted from a speaker discriminative deep neural network (DNN) yield better performance than the conventional methods such as i-vector.
  In most cases, the DNN speaker classifier is trained using cross entropy loss with softmax.
  However, this kind of loss function does not explicitly encourage inter-class separability and intra-class compactness. As a result, the embeddings are not optimal for speaker recognition tasks.
  In this paper, to address this issue, three different margin based losses which not only separate classes but also demand a fixed margin between classes are introduced to deep speaker embedding learning. It could be demonstrated that the margin is the key to obtain more discriminative speaker embeddings. Experiments are conducted on two public text independent tasks: VoxCeleb1 and Speaker in The Wild (SITW).
  The proposed approach can achieve the state-of-the-art performance, with 25\% $\sim$ 30\% equal error rate (EER) reduction on both tasks when compared to strong baselines using cross entropy loss with softmax, obtaining 2.238\% EER on VoxCeleb1 test set and 2.761\% EER on SITW core-core test set, respectively.

\end{abstract}

\section{Introduction}
Speaker recognition is the process of identifying or confirming the identity of a person given his speech segments.
Conventional speaker recognition system contains two phases: the first one is referred to as the enrollment phase, where the registered speaker's voices are converted to reference speaker model, while the second one is  the testing phase, where the recognition decision is made based on the enrolled model and the input testing speech.
If the speech content used in both stages is required to be the same, then the speaker recognition task is text-dependent, otherwise, it is text independent.

In current speaker recognition systems, using a low-dimensional fixed length vector, or speaker embedding, has become the dominant speaker modeling approach. 
Over the years, combined with probabilistic linear discriminative analysis (PLDA), i-vector~\cite{dehak2011front} has been the state-of-the-art system for text-independent speaker recognition.
Recently, with the development of deep neural networks and inspired by the great success of incorporating deep neural networks into speech recognition, researchers in the speaker recognition community also investigated the application of DNN for speaker modeling.
Furthermore, it is demonstrated that a high-performance speaker recognition system can be directly built from training a DNN speak classifier and extracting embeddings from it.
An utterance level DNN speaker embedding, named x-vector~\cite{snyder2016deep, snyder2017deep, snyder2018x}, that produced by a speaker discriminative DNN, has shown better performance than i-vector on a series of speaker recognition tasks.

However, the most widely used loss function for training the speaker-discriminative DNN is cross entropy loss with softmax (denoted by Softmax loss), which does not explicitly encourage inter-class separability and intra-class compactness.
While in speaker recognition, it is crucial that embeddings from the same identity aggregate and the clusters of different identities are well separated.
As a result, the embeddings produced by the DNN are not generalizable enough and performance degradation is observed when evaluated on unseen speakers.
Although entirely end to end system can do discriminative embedding learning directly~\cite{heigold2016end,zhang2017end,zhang2016end,huang2018joint}, it requires complicated data preparation such as semi-hard example mining and needs much longer time to train, thus it is not discussed in this work.

In this work, to encourage discriminative embedding learning, three losses that impose a fixed margin between classes are studied: angular softmax loss (denoted by A-Softmax loss)~\cite{liu2017sphereface}, additive margin softmax loss (denoted by AM-Softmax loss)~\cite{wang2018cosface,wang2018additive} and additive angular margin loss (denoted by AAM-Softmax loss)~\cite{deng2018arcface}.
It is found that the margin plays a vital role in learning discriminative embeddings and leads to a significant performance boost.
Experiments are conducted on two publicly available text independent tasks, VoxCeleb1 and Speaker in The Wild (SITW). When compared to strong baselines using Softmax loss, the proposed system can achieve 25\% $\sim$ 30\% EER reduction, with 2.238\% EER on VoxCeleb1 test set and 2.761\% EER on SITW core-core test set, respectively.

The rest of this paper is structured as follows. Section 2 gives a brief introduction of DNN speaker embedding systems.
Section 3 introduces three losses: A-Softmax, AM-Softmax and AAM Softmax, and compare the differences in designing the margin among them.
The experiment on VoxCeleb1 and SITW dataset are presented and analyzed in section 4.
Finally, section 5 concludes the paper.


\section{DNN speaker embedding systems}
The DNN speaker systems used in this work are based on the x-vector system described in~\cite{snyder2017deep,snyder2018x}, and most of them share the same configurations used in Kaldi's recipes.
All systems are built using Kaldi\cite{povey2011kaldi} or PyTorch~\cite{paszke2017automatic}.

The DNN architecture has five time-delay layers to handle the input at the frame level, followed by a statistical pooling layer that computes the mean and standard deviation of the input sequence, which aggregates the frame-level input into a segment-level representation. The following two dense layers operate on the segment-level input, and no temporal contexts are added.
Then, the projection layer maps the input to an output with dimension N, which is the total number of speakers in the training set. Rectified linear unit (ReLU) is used as the nonlinear transform and batch normalization (BN) is applied in all layers except for the projection layer.

The whole neural network is optimized using stochastic gradient descent. Speaker embeddings will be extracted from the segment-level layer of the well-trained DNN. 
However, in the literature~\cite{cai2018exploring,huang2018angular}, the effectiveness of Softmax loss for deep speaker embedding learning is questioned, which indicates a loss function which explicitly models the classification margins is promising.

\section{Losses for training the speaker discriminative DNN}
To study the effectiveness of adding margin when classifying different classes, the Softmax loss and three new loss functions (A-Softmax, AM-Softmax and AAM-Softmax) are first introduced.

\subsection{Softmax loss}
As a common used classification loss for training the speaker discriminative DNNs, Softmax loss can be formulated as:
$$L_\text{softmax}=-\frac{1}{N}\sum_{i=1}^N\log \frac{e^{\mathbf{W}^T_{y_i}\mathbf{x}_i+\mathbf{b}_{y_i}}}{\sum_{j=1}^c e^{\mathbf{W}^T_{j}\mathbf{x}_i+\mathbf{b}_{j}}}$$
where $N$ is the batch size, $c$ is the number of classes. $\mathbf{x}_i\in \mathbb{R}^d$ denotes the $i$-th input of samples to the projection layer and $y_i$ is the corresponding label index.
$\mathbf{W}\in \mathbb{R}^{d\times c}$ and $\mathbf{b}\in \mathbb{R}^{c}$ are the weight matrix and bias in the projection layer.

Since Softmax loss only penalizes on classification error, it does not explicitly enforce the similarity for intra-class samples and the diversity for inter-class samples.
This leads to a performance gap for speaker recognition tasks, which can be shown in the following experiments.

\subsection{A-Softmax loss}
In the definition of Softmax loss, if only the directions of the columns of the weight matrix $\mathbf{W}$ is considered and the bias term is discarded, the modified Softmax loss can be rewritten as

$$L_\text{Softmax}'=-\frac{1}{N}\sum_{i=1}^N\log \frac{e^{||\mathbf{x}_i||\cos(\theta_{{y_i}, i})}}{\sum_{j=1}^c e^{||\mathbf{x}_i||\cos(\theta_{{j}, i})}}$$

where $\theta_{j, i}$ is the angle between the column vector $\mathbf{W}_j$ and $\mathbf{x}_i$.
Then the embeddings learned by the modified Softmax loss have intrinsic angular distribution~\cite{liu2017sphereface,wen2016discriminative} since the probability of $\mathbf{x}_i$ belonging to class $j$ is only determined by the angle $\theta_{j, i}$.

Inspired by this form, a multiplicative angular margin can be incorporated into the modified Softmax loss, then the resulting A-Softmax loss can be defined as
$$L_\text{A-Softmax}=-\frac{1}{N}\sum_{i=1}^N\log \frac{e^{||\mathbf{x}_i||\phi(\theta_{{y_i}, i})}}{Z}$$
where 
$Z={e^{||\mathbf{x}_i||\phi(\theta_{{y_i}, i})} + \sum_{j=1,j\neq i}^c e^{||\mathbf{x}_i||\cos(\theta_{{j}, i})}}$
and $m\geq 1$ is the integer that controls the size of angular margin and  $$\phi(\theta_{{y_i}, i})=\cos(m\theta_{y_i, i})\leq \cos(\theta_{y_i, i})$$
for $\theta_{y_i, j}\in[0,\frac{\pi}{m}]$.

To remove the restriction of $\theta_{y_i, j}\in[0,\frac{\pi}{m}]$, function $\phi(\theta_{{y_i}, i})$ can be redefined as a piecewise monotonic decreasing function

$$\phi(\theta_{{y_i}, i})=(-1)^k\cos(m\theta_{y_i, i})-2k$$ for $k\in[0, m-1]$ and $\theta_{{y_i}, i}\in[\frac{k\pi}{m}, \frac{(k+1)\pi}{m}]$.

\subsection{AM-Softmax loss}
Although A-Softmax loss can impose a margin between classes, the piecewise function $\phi(\theta_{{y_i}, i})$ is not intuitive and optimization friendly.
Note that the margin can be imposed by design a function
$$\phi(\theta_{{y_i}, i})\leq \cos(\theta_{y_i, i})$$
An additive margin can be set instead of a multiplicative one and the definition of $\phi(\theta_{{y_i}, i})$ is 
$$\phi(\theta_{{y_i}, i})=\cos(\theta_{y_i, i}) - m$$

Moreover, in the definition of A-Softmax loss, only the columns of the weight matrix are normalized and no restrictions are applied to the length of input vector, which may be harmful to learning the distances between embeddings, since the distance between two points come from different classes can be very small if they are both close to the origin.


As a result, the input $\mathbf{x}_i$ is normalized as well, and the additive margin softmax loss is defined as
$$L_\text{AM-Softmax}=-\frac{1}{N}\sum_{i=1}^N\log \frac{e^{s({\cos(\theta_{{y_i}, i})-m})}}{Z}$$
where
$Z={e^{s({\cos(\theta_{{y_i}, i})-m)}} + \sum_{j=1,j\neq i}^c e^{s({\cos(\theta_{{j}, i})})}}$
and $s$ is a scaling factor used to make sure the gradient not too small during the training~\cite{wang2017normface}.

\subsection{AAM-Softmax loss}
In AM-Softmax loss, $\mathbf{x}_i$ are normalized into unit vectors before the projection layer, which means $\mathbf{x}_i$ are points in a hypersphere.
Then the arc connects $\mathbf{x}_i$ and $\mathbf{x}_j$ can give a natural definition of the distance between them: the arc's length.
While the arc's length exactly corresponds to the angle between the unit vectors $\mathbf{x}_i$ and $\mathbf{x}_j$, a more natural and intuitive definition of $\phi(\theta_{{y_i}, i})$
is 
$$\phi(\theta_{{y_i}, i})=\cos(\theta_{y_i, i} + m)$$
This leads to the definition of additive angular margin loss:
$$L_\text{AAM-Softmax}=-\frac{1}{N}\sum_{i=1}^N\log \frac{e^{s({\cos(\theta_{{y_i}, i}+m)})}}{Z}$$
where $Z={e^{s({\cos(\theta_{{y_i}, i}+m))}} + \sum_{j=1,j\neq i}^c e^{s({\cos(\theta_{{j}, i})})}}$.

\section{Experiments}
The experiments are carried out on the VoxCeleb1~\cite{nagrani2017voxceleb} and SITW~\cite{mclaren2015speakers} test set with most settings are the same, while detailed training data differ, which will be introduced in the corresponding sections.

\subsection{Basic experimental set-up}

\subsubsection{Data preparation}
To increase the amount and diversity of the training data, the same kind of data augmentation in~\cite{snyder2017deep,snyder2018x} is applied to add noises, music, babble and reverberation. 

The features are 30-dimensional Mel-Frequency Cepstral Coefficients (MFCCs), with a frame shift of 10ms and a window width of 25ms. Mean normalization is then applied over a sliding window of up to 3 seconds.
To filter out non-speech frames, an energy-based voice activity detector (VAD) is employed.

\subsubsection{Architecture}
The architecture of the speaker discriminative DNN used in this work is illustrated in Table~\ref{tab:arch}, which is similar to the one used in Kaldi~\cite{povey2011kaldi}'s recipes (v2) for VoxCeleb1 or SITW. The width of the projection layer's output varies according to the different number of speakers in the training set. After training, the 512-dimensional speaker embeddings are extracted from segment6's affine layer given the input features.

\begin{table}[th]
    \caption{Architecture of the speaker discriminative DNN}
    \label{tab:arch}
    \centering
    \begin{tabular}{@{}cccc@{}}
        \toprule
        \multicolumn{1}{c}{\textbf{Layer}} & 
        \multicolumn{1}{c}{\textbf{Layer context}} &
        \multicolumn{1}{c}{\textbf{Total context}} &
        \multicolumn{1}{c}{\textbf{Input$\times$output}}  \\
        \midrule
        frame1 & [t-2, t+2] & 5 & 150$\times$512 \\
        frame2 & \{t-2, t, t+2\} & 9 & 1536$\times$512 \\
        frame3 & \{t-3, t, t+3\} & 15 & 1536$\times$512 \\
        frame4 & \{t\} & 15 & 512$\times$512 \\
        frame5 & \{t\} & 15 & 512$\times$1500 \\
        stats pool & [0, T-1) & T & 1500T$\times$3000 \\
        segment6 & \{0\} & T & 3000$\times$512 \\
        segment7 & \{0\} & T & 512$\times$512 \\
        projection & \{0\} & T & 512$\times$N \\
        \bottomrule
    \end{tabular}
    
\end{table}

\begin{table*}[h]
    \caption{Results on the original VoxCeleb1 test set and the extented and hard test sets (VoxCeleb1-E and VoxCeleb1-H).}
    \label{tab:res2}
    \centering
    \begin{tabular}{ccccc}
        \toprule
         & \textbf{Model} & \textbf{Loss} & \textbf{Training set} & \textbf{EER} \\
        \midrule
        \textbf{VoxCeleb1 test set} & & & & \\
        \midrule
        Nagrani {\it et al.}~\cite{nagrani2017voxceleb} & GMM-UBM (i-vector) & - & VoxCeleb1 & 8.8 \\
        Cai {\it et al.}~\cite{cai2018exploring} & ResNet-34 & A-Softmax & VoxCeleb1 & 4.40 \\
        Okabe {\it et al.}~\cite{okabe2018attentive} & TDNN (x-vector) & Softmax & VoxCeleb1 & 3.85  \\
        Hajibabaei {\it et al.}~\cite{hajibabaei2018unified} & ResNet-20 & AM-Softmax & VoxCeleb1 & 4.30  \\
        Chung {\it et al.}~\cite{Chung18b} & ResNet-50 & Softmax + Contrastive & VoxCeleb2 & 4.19 \\
        Xie {\it et al.}~\cite{xie2019utterance} & Thin ResNet-34 & Softmax & VoxCeleb2 & 3.22 \\
        \textbf{Ours} & TDNN (x-vector) & AAM-softmax & VoxCeleb2 & \textbf{2.694} \\
        \textbf{Ours} & TDNN (x-vector) & AAM-softmax & VoxCeleb1 + VoxCeleb2 & \textbf{2.238} \\
        \midrule
        \textbf{VoxCeleb1-E test set} & & & & \\
        \midrule
        Chung {\it et al.}~\cite{Chung18b} & ResNet-50 & Softmax + Contrastive & VoxCeleb2 & 4.42 \\
        Xie {\it et al.}~\cite{xie2019utterance} & Thin ResNet-34 & Softmax & VoxCeleb2 & 3.13 \\
        \textbf{Ours} & TDNN (x-vector) & AAM-softmax & VoxCeleb2 & \textbf{2.762} \\
        \midrule
        \textbf{VoxCeleb1-H test set} & & & & \\
        \midrule
        Chung {\it et al.}~\cite{Chung18b} & ResNet-50 & Softmax + Contrastive & VoxCeleb2 & 7.33 \\
        Xie {\it et al.}~\cite{xie2019utterance} & Thin ResNet-34 & Softmax & VoxCeleb2 & 5.06 \\
        \textbf{Ours} & TDNN (x-vector) & AAM-softmax & VoxCeleb2 & \textbf{4.732} \\
        \bottomrule
    \end{tabular}
\end{table*}

\subsubsection{Training}
Similar to the strategy introduced in \cite{snyder2018x}, the systems are trained on segments ranges from $2\sim 4$ seconds, which are obtained by randomly cutting the original utterances. The training of all DNNs are done with PyTorch.
Stochastic gradient descent (SGD) with the moment is used to optimize the DNN. 
To reduce the training time, Horovod~\cite{sergeev2018horovod} is facilitated to coordinate the synchronous SGD training over 8 GeForce GTX 1080Ti GPUs, with the batch size of 64 on each GPU.
Each model's parameters on 8 GPUs are initialized with the same random seed.
At each training step, the gradients on all GPUs are accumulated and then sent back to each GPU by ring allreduce algorithm to reduce the communication overhead.

To make the training at the beginning more stable, as suggested in~\cite{goyal2017accurate} the learning rate is set to $0$ and then gradually increased to $1\text{e-}4$ at the first 65,536 batches (8,192 batches per GPU).

All systems are trained for 3 epochs with a learning rate $1\text{e-}4$, momentum $0.7$, weight decay $1\text{e-}5$ and a maximum gradient norm $1\text{e}3$.

\subsubsection{Scoring}
Among various variants of PLDA models, the standard version introduced in \cite{ioffe2006probabilistic} and implemented in Kaldi \cite{povey2011kaldi} is used as the scoring back-end for all the systems.
First the embeddings are centered and projected to 128-dimensional representations using linear discriminative analysis (LDA), then the representations are length normalized and modeled by PLDA.

\subsection{System evaluation on VoxCeleb1}

\subsubsection{Training data}
The speaker discriminative DNN model is trained on all of VoxCeleb2~\cite{Chung18b} plus the training portion of VoxCeleb1, which are sampled at 16kHz.
This leaves a total of 1,277,503 utterances from 7,325 speakers.

A random subset consists of 1,000,000 utterances of the augmentations is kept and combined with the original training data.
As a result, the final training data consists of 2,128,429 utterances after silence removal.

\subsubsection{Evaluation}
There are four kinds of systems with different losses evaluated on the test set of VoxCeleb1.
For the two hyperparameters in the definition of losses, the scale $s$ is fixed to $32$ in all systems, while the margin $m$ is tested with two values for each system, and only the one with better performance is presented.
Equal error rate (EER) and minimum detection cost function (minDCF) with p-target of 0.01 or 0.001 are used as performance metrics.

The results of four systems equipped with different kinds of losses are presented in Table~\ref{tab:res1}.
The first two lines report the number of two baseline systems that are trained by Kaldi and Pytorch respectively. Both systems give similar results, which confirms the correctness of the implementations trained with PyTorch.

The next three lines report the results of proposed systems by replacing standard Softmax loss to A-Softmax loss, AM-Softmax loss and AAM-Softmax loss, respectively.
Compared with the baseline system, the three proposed systems can outperform it by a large margin.
Especially for the AAM-Softmax system, which achieves an EER of 2.238\%, or equivalently a 30\% reduction in EER compared to the baseline numbers.
To the best of our knowledge, this is the best number published on VoxCeleb1 test set.
All these results confirm the importance of incorporating the margin in embedding learning.

\begin{table}[th]
    \caption{Comparison of systems under the VoxCeleb1 test set. All systems are trained on VoxCeleb1 trainining set and the whole VoxCeleb2 set with data augmentation.}
    \label{tab:res1}
    \centering
    \begin{tabular}{@{}ccccc@{}}
        \toprule
        \textbf{System} & \textbf{m} & \textbf{EER} & \textbf{minDCF$_{0.01}$} & \textbf{minDCF$_{0.001}$} \\
        \midrule
        Softmax (Kaldi) & - & 3.208 & 0.3481 & 0.5753 \\
        Softmax & - & 3.271 & 0.3646 & 0.5018 \\
        \midrule
        A-Softmax & 2 & 2.434 & 0.2774 & 0.4536 \\
        AM-Softmax & 0.2 & 2.264 & 0.2537 & \bf{0.3293} \\
        AAM-Softmax & 0.3 & \bf{2.238} & \bf{0.2433} & 0.4119 \\
        \bottomrule
    \end{tabular}
\end{table}

\subsubsection{Using VoxCeleb2 development set only}
This section presents the performance of systems trained only on VoxCeleb2 development set, which is fairer when comparing to other state-of-the-art systems.
The training data consists of 5,994 speakers and is entirely disjoint from the VoxCeleb1 dataset, and no data augmentation is used.
Under the settings, two extra test sets are used for evaluation: the extended VoxCeleb1-E that uses the entire VoxCeleb1 (train and test splits) and the challenging VoxCeleb1-H that the test pairs are drawn from identities with the same gender and nationality.

Table~\ref{tab:res2} compares the performance of the proposed models to state-of-the-art on three test sets: VoxCeleb1, VoxCeleb1-E and VoxCeleb1-H.
On all three test sets, the best proposed systems can achieve 16\%, 12\% and 6\% reduction in EER when compared with the previous state-of-the-art.

\subsection{System evaluation on SITW}
\subsubsection{Training data}
The speaker discriminative DNN model is trained on the development portion of VoxCeleb2 plus the whole set of VoxCeleb1, which are sampled at 16kHz.
The test portion of VoxCeleb2 is not used since it has overlapped speakers in the SITW test set. 
Besides, there are 60 speakers in VoxCeleb1 that overlap with the SITW core-core test set and they are removed.
This leaves a total of 1,236,567 utterances from 7,185 speakers.
The same kind of data augmentation used in the experiments on VoxCeleb1 test set is applied.
A random subset of 1,000,000 utterances of the augmentations is kept and combined with the original training data.
Finally, the final training data consists of 2,090,306 utterances after silence removal.

\subsubsection{Evaluation}
As shown in Table~\ref{tab:res3}, the proposed systems outperform the baselines a lot and the reduction in EER is 20\%, 17\% and 25\% when using A-Softmax loss, AM-Softmax loss and AAM-Softmax loss respectively.
Since the utterances from the enrollment and test set of SITW vary in length from $6\sim 240$ seconds and extra efforts for compensation are necessary, the results may not fully reflect the gains from using the proposed loss functions.

\begin{table}[th]
    \caption{Comparison of systems under the SITW test set. All systems are trained on the whole VoxCeleb1 set and VoxCeleb2 development set with data augmentation. 60 speakers in VoxCeleb1 that overlap with the test set are removed.}
    \label{tab:res3}
    \centering
    \begin{tabular}{@{}ccccc@{}}
        \toprule
        \textbf{System} & \textbf{m} & \textbf{EER} & \textbf{minDCF$_{0.01}$} & \textbf{minDCF$_{0.001}$} \\
        \midrule
        Softmax (Kaldi) & - & 3.581 & 0.3456 & 0.5165 \\
        Softmax & - & 3.718 & 0.3491 & 0.5195 \\
        \midrule
        A-Softmax & 2 & 2.980 & 0.3045 & 0.5048 \\
        AM-Softmax & 0.2 & 3.089 & \textbf{0.2931} & \textbf{0.4496} \\
        AAM-Softmax & 0.2 & \textbf{2.761} & 0.3002 & 0.4712 \\
        \bottomrule
    \end{tabular}
\end{table}

\section{Conclusions}
Good speaker embeddings are expected to have large inter-speaker difference while retaining small intra-speaker variation, which needs strong discrimination supervision signals from the training criterion to guarantee such properties. Most current deep speaker embedding frameworks utilize the Softmax loss as the optimization criterion, which is proved inferior to the more advanced margin-based classification loss functions. In this paper, three margin-based loss functions, {\it i.e.}, A-Softmax, AM-Softmax and AAM-Softmax, are introduced to the x-vector based speaker embedding learning framework. The proposed systems are evaluated on two test sets: VoxCeleb1 and SITW and the results show that proposed methods significantly outperform the baseline.

The best proposed system achieves 25\% $\sim$ 30\% equal error rate (EER) reduction on both tasks when compared to strong baselines using cross entropy loss with softmax, obtaining 2.238\% EER on VoxCeleb1 test set and 2.761\% EER on SITW core-core test set, respectively, which represents the state-of-the-art performance.
\bibliographystyle{IEEEtran}

\bibliography{template}


\end{document}